\documentstyle[aps,prl,floats]{revtex}
\input{epsf}
\begin{document}
\draft

\twocolumn[\hsize\textwidth\columnwidth\hsize\csname@twocolumnfalse\endcsname
\title{Using mean field theory to determine the structure of uniform fluids}
\author{Katharina Vollmayr-Lee$^1$, Kirill Katsov$^2$, and John
D. Weeks$^{2,3}$}
\address{$^1$Department of Physics, Bucknell University, 
Lewisburg, Pennsylvania 17837\\$^2$Institute for Physical
Science and Technology and $^3$Department of Chemistry and Biochemistry,\\
University of Maryland, College Park, Maryland 20742}
\date{\today }
\maketitle

\begin{abstract}
The structure of a uniform simple liquid is related to
that of a reference fluid with purely repulsive intermolecular forces
in a self-consistently determined external reference field (ERF)
$\phi _ R$. The ERF can be separated into a harshly repulsive
part  $\phi _ {R0}$ generated by the repulsive core of a reference
particle fixed at the origin and a more slowly varying part
$\phi _{R1}$ arising from a mean field treatment
of the attractive forces.  We use a generalized linear
response method to calculate the reference fluid structure,
first determining the response to the smoother part $\phi _{R1}$  of the ERF
alone,  followed by the response to the harshly repulsive part.
Both steps can be carried out very accurately, as confirmed by MD
simulations, and good agreement with the structure of the full LJ fluid
is found.
\end{abstract}

\pacs{05.20.Jj, 61.20.Gy, 61.20.Ne}
]

\section{Introduction}

\label{Intro} In this paper we describe a new and physically motivated way
to determine the structure of uniform fluids, based on a mean field
treatment of the attractive intermolecular interactions. To apply this
approximation to a general nonuniform system, attractive interactions are
replaced by a spatially varying single particle ``molecular field''
potential, chosen to take account of variations in the average attractive
energy density in different parts of the system \cite{rolwidom}. Since the
attractive interactions usually operate over an extended range, it seems
plausible that an averaged description as given by mean field theory could
often provide a useful simplification.

However, in real liquids, additional very important ``excluded volume''
correlations are generated by the short-ranged and harshly repulsive
intermolecular forces \cite{hansenmac,wca}; these cannot be accurately
described using the same mean field averaging appropriate for the longer
ranged attractive forces. Despite this additional complexity, the use of
mean field theory allows us to consider an inherently simpler system: a {\em %
nonuniform reference fluid}. This consists of particles interacting only
with repulsive intermolecular forces but in the presence of an {\em external
reference field} (ERF) chosen self-consistently to take account of the
locally averaged effects of attractive interactions as well as any imposed
external field \cite{wsb,wvk,wkv}. The uniform reference fluid is stable
over the entire range of densities from vapor to liquid, and its structure
in the presence of an appropriately chosen ERF approximates that of the
original system.

In previous work \cite{wsb,wvk,wkv} we showed how these ideas can be used to
give an accurate description of the structure of a nonuniform Lennard-Jones
(LJ) fluid in a number of different applications where the ERF is large and
attractive forces strongly influence the structure, including the
liquid-vapor interface and the structure of fluids near hard walls. In these
cases conventional (singlet) integral equation methods have given poor
results \cite{sullstell}.

In this paper we consider a different limit, that of the {\em uniform} LJ
fluid. Here attractive forces produce relatively small structural changes at
high density and integral equation methods have had their greatest
successes. Indeed, in the simplest picture, the attractive forces on a given
particle from oppositely situated neighbors essentially {\em cancel} \cite
{widomsci} in typical high density configurations, and the structure of the
dense uniform LJ fluid is rather well approximated by that of a uniform
reference fluid at the same density \cite{wca}. The theoretical challenge is
to improve on this rather accurate starting point at high density and to
describe the larger structural changes attractive interactions induce at
lower densities.

From this perspective the uniform LJ fluid provides an important and
nontrivial test of our general approach. It is not clear that mean field
averaging along with the approximate methods \cite{wkv,kw} we use to
calculate reference fluid structure will be accurate enough to determine the
small but subtle changes induced by attractive interactions in the highly
oscillatory structure of uniform fluids at high density or the more
substantial changes seen at lower densities. Indeed, unlike the previous
applications, it is difficult to guess even qualitative features of the ERF.

The plan of the paper is the following. In Sec.~II we define the nonuniform
reference fluid and the formal equation determining the ERF. In Sec.~III we
discuss the usual mean field approximation for the ERF and suggest a new
generalized equation that incorporates exact results at low density.

In Sec.~IV we first use computer simulations to carry out the determination
of structure essentially exactly. This allows us to test the accuracy of the
basic mean field description of the ERF without any further approximations.
We generally find quite satisfactory results though small errors in the
simplest mean field determination of the ERF can be seen at high density.

In Sec.~V we introduce a new theory to calculate self-consistently both the
ERF and the associated structure in the reference fluid, using a
generalization of a physically motivated procedure first used to calculate
the structure of a LJ\ fluid near a hard wall \cite{wkv}. The key idea is to
divide the ERF into rapidly and slowly varying parts. We determine the
response of the reference fluid density to each component of the ERF in
successive steps, using appropriate methods in each step that can accurately
describe the very different density responses, as discussed in Secs.~\ref
{firststep} and \ref{secondstep}. Sec.~\ref{results} discusses the results
of the method and comparison to simulations. The theory generally gives
quite satisfactory results. However, at the highest densities some small
errors can be seen arising from the simple mean field treatment of
attractive interactions. At lower densities, our simplest approximations
give results comparable to the best integral equation methods \cite
{lado73,hmsa} Final remarks are given in Sec.~\ref{Finalremarks}. Some
technical details are presented in the Appendices.

\section{Nonuniform Reference Fluid}

We first consider the case where fluid particles interact with a known
external field $\phi ({\bf r})$ and through the LJ pair potential $%
w(r)\equiv u_{0}(r)+u_{1}(r)$, separated as usual \cite{wca} into rapidly
and slowly varying parts so that all the repulsive intermolecular forces
arise from $u_{0}$ and all the attractive forces from $u_{1}$ \cite{cutoff}.
We assume that the external field $\phi ({\bf r})\equiv $ $\phi _{0}({\bf r}
)+\phi _{1}({\bf r})$ can be separated in a similar way, where the subscript 
$0$ denotes a harshly repulsive interaction and the subscript $1$ a much
more slowly varying interaction usually associated with attractive forces.
We consider a grand ensemble with fixed chemical potential $\mu ^{B}$, which
determines $\rho ^{B}$, the uniform fluid density when $\phi =0$.

We relate the structure of the nonuniform LJ\ system to that of a simpler 
{\em nonuniform} {\em reference fluid }\cite{wsb,wvk,wkv}, with only
repulsive intermolecular pair interactions $u_{0}(r_{ij})$ (equal to the LJ
repulsions) in a different {\em effective reference field} (ERF) $\phi _{R}(%
{\bf r})$. The replacement of attractive pair interactions by an approximate
local ``molecular field'' is an essential step in mean field theory, but we
can think of other more general prescriptions for $\phi _{R}({\bf r})$. Here
we determine $\phi _{R}({\bf r})$ formally by the requirement that it has a
functional form such that the {\em local} (singlet) density at every point $%
{\bf r}$ in the reference fluid equals that of the full LJ fluid \cite
{sullstell}: 
\begin{equation}
\rho _{0}({\bf r;[}\phi _{R}])=\rho ({\bf r;[}\phi ])\,.  \label{singletden}
\end{equation}

The subscript $0$ in Eq.~(\ref{singletden}) reminds us that the reference
system pair interactions arise only from $u_{0}$ and the notation ${\bf [}%
\phi _{R}]$ indicates that all distribution functions are functionals of the
appropriate external field. Since $\rho ({\bf r;[}\phi ])\,$is a physically
realizable distribution function, and the reference fluid is stable over a
wide range of densities, it seems very plausible that such a choice for the
field $\phi _{R}$ can be made in principle \cite{lit:denfungeneral}. In
practice we will make approximate choices motivated by mean field ideas.

Using this perspective, let us consider the problem of determining the
effects of attractive forces on the structure of {\em uniform} fluids. The
simplest (WCA) approximation \cite{wca} assumes complete cancellation of
attractive forces and approximates the radial distribution function $g(r)$
of the uniform LJ fluid by the $g_{0}(r)$ of the uniform repulsive force
fluid at the same density. To improve on this, we make use of the exact
relation \cite{percus62} between $\rho ^{B}g(r)$ in the uniform LJ fluid and the
singlet density in a nonuniform fluid with a particle fixed at ${\bf r}_{0}$,
which we take to be the origin of our coordinate system: 
\begin{equation}
\rho ^{B}g(r_{1})=\rho ({\bf r}_{1}|{\bf r}_{0};[\phi =0])\,\,=\,\rho
(r_{1};[\phi _{LJ}])\,.  \label{origin}
\end{equation}
Here $\rho ({\bf r}_{1}|{\bf r}_{0};[\phi =0])$ is the {\em conditional}
singlet density --- the density in zero external field at ${\bf r}_{1}$
given that a particle is fixed at ${\bf r}_{0}$. By symmetry this depends
only on the radial distance $r_{1}\equiv |{\bf r}_{1}|$ from the fixed
``wall particle'' at ${\bf r}_{0}=0.$ This in turn must equal the nonuniform
singlet density induced by the external field $\phi _{LJ}(r_{1})\equiv
w(r_{1})$.

By choosing $\phi _{R}(r_{1})$ in Eq.~(\ref{singletden}) to fit the
nonuniform LJ density $\rho (r_{1};[\phi _{LJ}])$, we obtain a {\em %
nonuniform reference system} in which the density $\rho _{0}(r_{1}{\bf ;[}%
\phi _{R}])$ is modified by the effects of attractive forces. In particular
this can be used in Eq.~(\ref{origin}) to calculate the radial distribution
function $g(r_{1})$ of the uniform LJ system. The original WCA approximation 
\cite{wca} arises from the particular choice $\phi _{R}=u_{0}.$ See Fig.~(1).

%%%%%%%%%%%%%
%%%%%%%%%%%%%
\begin{figure}[tbp]
\epsfxsize=3.2in
\epsfbox{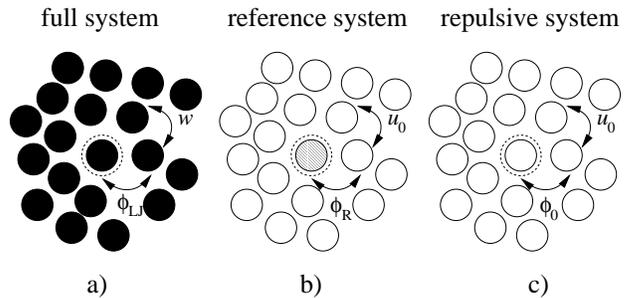}
\caption{Model systems considered. a) is the full LJ system with a LJ particle
fixed at the origin as indicated by the dashed circle. The interaction with
the other particles can be described by an external field
$\phi _{LJ}(r)=w(r).$ b) is the nonuniform reference system with the special
wall particle fixed at the origin with interaction $\phi _{R}(r).$ c) is the
original WCA repulsive force system with pair potential $u_{0}(r)$. Here the
fixed particle interacts with the other particles through
$\phi _{0}(r)=u_{0}(r).$} 
\label{fig1}
\end{figure}
%%%%%%%%%%%%%
%%%%%%%%%%%%%

\section{Mean field approximation for the ERF}

\label{meanfielddeter} In previous work \cite{wsb,wvk,wkv} we started from
the balance of forces as described by the exact Yvon-Born-Green hierarchy 
\cite{hansenmac} and arrived at a generalized mean field equation for the
ERF by a series of physically motivated approximations. We will not repeat
these arguments here and instead focus on the simplest final approximation,
the {\em molecular field equation} for the ERF, which proved surprisingly
accurate in a number of different applications. This is just a transcription
of the usual molecular field equation for the Ising model \cite{rolwidom}
to a continuum fluid with attractive interactions $u_{1}(r)$ and can be
immediately written down: 
\begin{equation}
\phi _{R}^{MF}(r_{1})=\phi _{LJ}(r_{1})+\int d{\bf r}_{2}\,[\rho _{0}(r_{2};%
{\bf [}\phi _{R}^{MF}])-\rho ^{B}]\,u_{1}(r_{12})\,.  \label{mfint}
\end{equation}
The effective field $\phi _{R}^{MF}$ at a particular distance $r_{1}$ from
the fixed wall particle is comprised of the bare field $\phi _{LJ}(r_{1})$
from the fixed particle plus the integral over all positions ${\bf r}_{2}$
of the attractive interactions $u_{1}(r_{12})$ from other particles weighted
by the deviation of the nonuniform reference density $\rho _{0}(r_{2};{\bf [}%
\phi _{R}^{MF}])$ from its limiting value $\rho ^{B}$. Use of the density
deviation ensures that $\phi _{R}^{MF}$ vanishes at large $r_{1}$.

Let $\phi _{s}$ denote the second term on the right in Eq.~(\ref{mfint}): 
\begin{equation}
\phi _{s}(r_{1})\equiv \int d{\bf r}_{2}\,[\rho _{0}(r_{2};{\bf [}\phi
_{R}^{MF}])-\rho ^{B}]\,u_{1}(r_{12})\,.  \label{phis}
\end{equation}
It provides an estimate of the averaged effects of attractive pair
interactions arising from the other (mobile) particles in the full LJ fluid
at a distance ${\bf r}_{1}$ from a particle fixed at the origin. Because of
the convolution with the slowly varying attractive potential ``weighting
function'' $u_{1}(r_{12})$ in Eq.~(\ref{phis}), $\phi _{s}(r_{1})$ extends
smoothly into the repulsive core region of the wall particle where $\rho
_{0}(r_{1};{\bf [}\phi _{R}^{MF}])$ vanishes. Outside the core it is a
smooth, basically repulsive and relatively slowly varying interaction even
when $\rho _{0}(r_{1};{\bf [}\phi _{R}^{MF}])$ itself has pronounced
oscillations.

More complicated, but sometimes more accurate, equations for the ERF are
available \cite{wvk}, but in practice the simple mean field approximation (%
\ref{mfint}) often gives quite satisfactory results. In Appendix A we
discuss a simple modification of Eq.~(\ref{phis}) that gives somewhat more
accurate results at low density. In the following we will use Eq.~(\ref
{mfint}) to determine the ERF unless otherwise indicated.

\section{Results from MD simulations}

We now must solve Eq.~(\ref{mfint}) to determine the ERF $\phi _{R}^{MF}$
and associated density $\rho _{0}(r;{\bf [}\phi _{R}^{MF}]).$ As is typical
in mean field theory, a self-consistent solution must be found, since the
ERF $\phi _{R}$ appears explicitly on the left side and implicitly on the
right side through the dependence of the density $\rho _{0}(r_{2};{\bf [}%
\phi _{R}])$ on $\phi _{R}$. If we can find the reference structure $\rho
_{0}(r;{\bf [}\phi _{R}])$ produced by a given ERF $\phi _{R}$ accurately,
then it is straightforward to solve the mean field equation (by iteration,
for example) to determine the self-consistent $\phi _{R}^{MF}$ and the
associated density $\rho _{0}(r;{\bf [}\phi _{R}^{MF}])$. In Sec.~\ref
{TwoStepSec} we will discuss new theoretical methods to calculate $\rho
_{0}(r;{\bf [}\phi _{R}])$ for a given $\phi _{R}$. However, since these
could introduce additional errors, it is useful first to assess the accuracy
of the basic mean field equation (\ref{mfint}) without any further
approximations.

To that end, we carried out MD simulations for the three model systems shown
in Fig.~(1): the full LJ system, the WCA repulsive force system, and the
inhomogeneous reference system with the special wall particle fixed at the
origin \cite{preliminary}. To determine the effective potential in the
latter case, we solved Eq.~(\ref{mfint}) by iteration \cite{iteration},
using (essentially exact) MD results for $\rho _{0}(r;{\bf [}\phi _{R}])$.
The errors in $\rho _{0}(r;{\bf [}\phi _{R}^{MF}])$ when compared to $\rho
(r;{\bf [}\phi _{LJ}])$ then arise solely from the mean field approximation
for the ERF $\phi _{R}^{MF}.$

\subsection{Simulation details}

In the following we use reduced Lennard-Jones units where the unit of length
is $\sigma $, the unit of energy is $\epsilon $ and the unit of time is $%
\sqrt{m\sigma ^{2}/\epsilon }.$ We carried out MD simulations in the
(NVT)-ensemble using the velocity Verlet algorithm with a time step of $%
\Delta t=0.001$. To maintain constant temperature, every 150 MD steps we
chose new velocities for all particles from the corresponding Boltzmann
distribution.

We simulated states along the near critical isotherm at $T=1.35$ for
densities $\rho ^{B}=0.78$, $0.54$, $0.45$ and $0.1,$ and a state near the
triple point with $T=0.88$ and $\rho ^{B}=0.85$. We used $N=3000$ particles
for $T=1.35,\rho ^{B}=0.78$ and for $T=0.88,\rho ^{B}=0.85,$ and $N=450$ for
all other states. To eliminate the possibility of finite size effects we
made test runs for $T=1.35$ and $\rho ^{B}=0.54,0.45$ and $0.1$ with $N=3000$
particles, which led to the same density distributions as for $N=450$. Each
state was first equilibrated for $5\cdot 10^{5}$ MD steps. Subsequently we
calculated $g(r)$ for the uniform systems and $\rho _{0}(r;[\phi _{R}^{MF}])$
for the nonuniform reference system for at least $3.5\cdot 10^{6}$ and up to 
$7.5\cdot 10^{7}$ MD steps.

\subsection{Simulation results for the ERF}

\label{simulationresults} We first concentrate on the high density state $%
\rho ^{B}=0.78$ and $T=1.35$, which will illustrate many basic features of
the mean field approach. Figure (2) gives simulation results for the full LJ
density $\rho ^{B}g(r),$ the WCA reference density $\rho ^{B}g_{0}(r),$ and
the inhomogeneous reference density $\rho _{0}(r;[\phi _{R}^{MF}])$ for this
state. We see that the mean field prediction $\rho _{0}(r;[\phi _{R}^{MF}])%
\approx \rho ^{B}g(r)$ is able to correct the main quantitative errors in
the already rather accurate WCA approximation $g_{0}(r)\approx g(r)$,
describing in particular the slight shift outward of the first peak.
However, some small errors remain, due solely to the mean field
approximation for the ERF from Eq.~(\ref{mfint}). These are focused on in
the inset to Fig.~(2), which compares the density change $\rho _{0}(r;[\phi
_{R}^{MF}])-\rho ^{B}g_{0}(r)$ due to attractive forces as predicted by Eq.~(%
\ref{mfint}) to the actual change $\rho ^{B}[g(r)-g_{0}(r)]$ given by the
simulations. Any further improvements in these results will require a better
approximation for the ERF.

%%%%%%%%%%%%%
%%%%%%%%%%%%%
\begin{figure}[tbp]
\epsfxsize=3.2in
\epsfbox{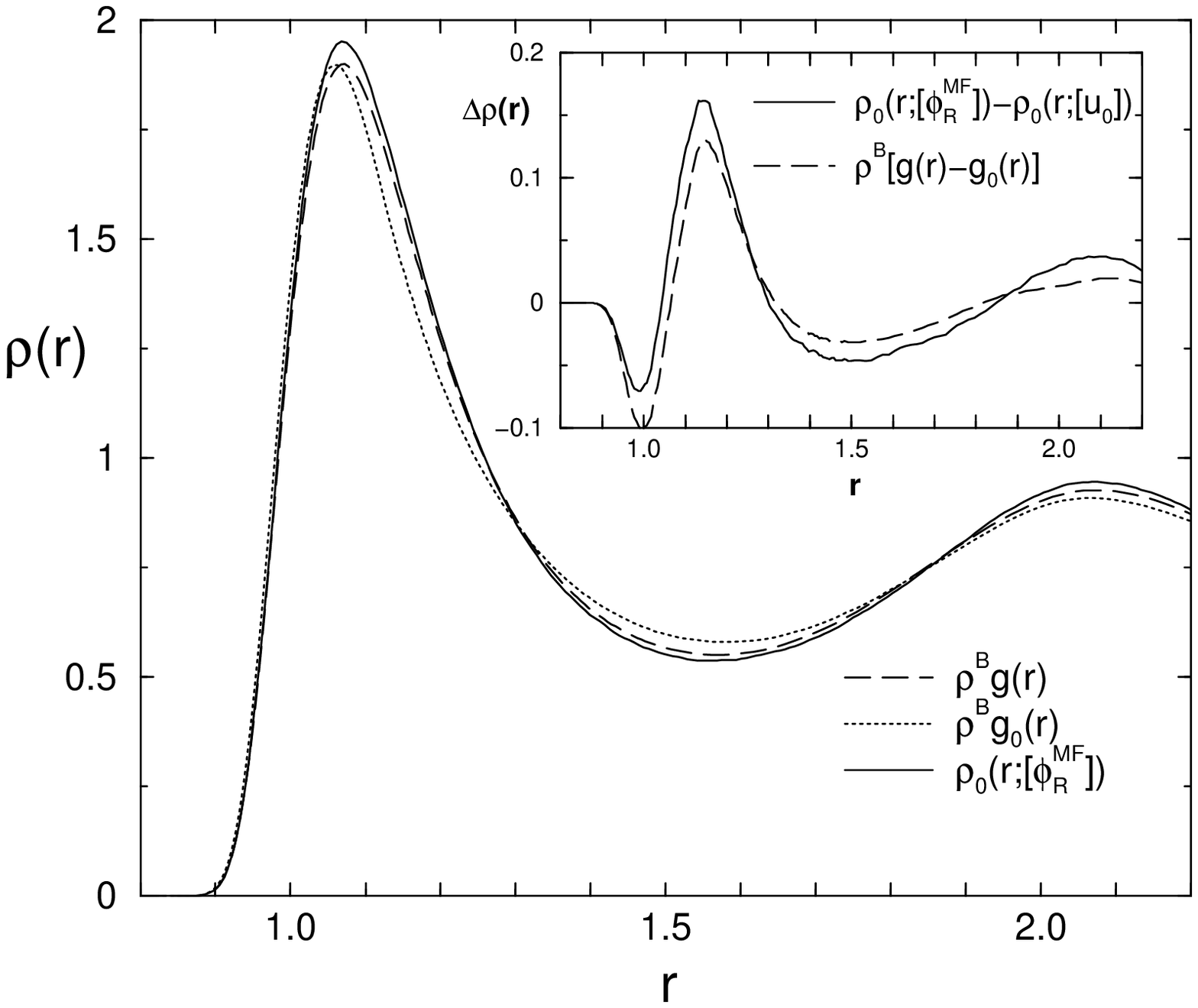}
\caption{Densities with a particle fixed at the origin as determined by MD
simulations for the three systems in Fig.~1 at $T=1.35$ and $\rho ^{B}=0.78.$
The inset gives the difference in density between the LJ fluid and the WCA
repulsive force fluid as determined by simulations and by self-consistent
solution of the mean field equation (\ref{mfint}), again using simulation
data.}
\label{fig2}
\end{figure}
%%%%%%%%%%%%%
%%%%%%%%%%%%%

Figure (3) shows the corresponding self-consistent ERF $\phi _{R}^{MF}(r)$
from Eq.~(\ref{mfint}), compared to the bare LJ potential $w(r)$ and the
repulsive reference potential $u_{0}(r).$ At low density $\phi _{R}^{MF}$
reduces exactly to $w$, and if the force cancellation argument were exact,
then at high density $\phi _{R}^{MF}$ would equal $u_{0}$ as assumed in the
WCA approximation. However, there is a weak negative region in $\phi
_{R}^{MF}(r)$ for $r$ between about $1.1$ and $1.4.$ This results from the
nonuniform attractive energy density experienced by a particle in this
region in the LJ system, which is slightly lower than average because of the
fixed particle and its neighbors even at this high density. This produces
the slight shift in the first peak noted above.

%%%%%%%%%%%%%
%%%%%%%%%%%%%
\begin{figure}[t]
\epsfxsize=3.2in
\epsfbox{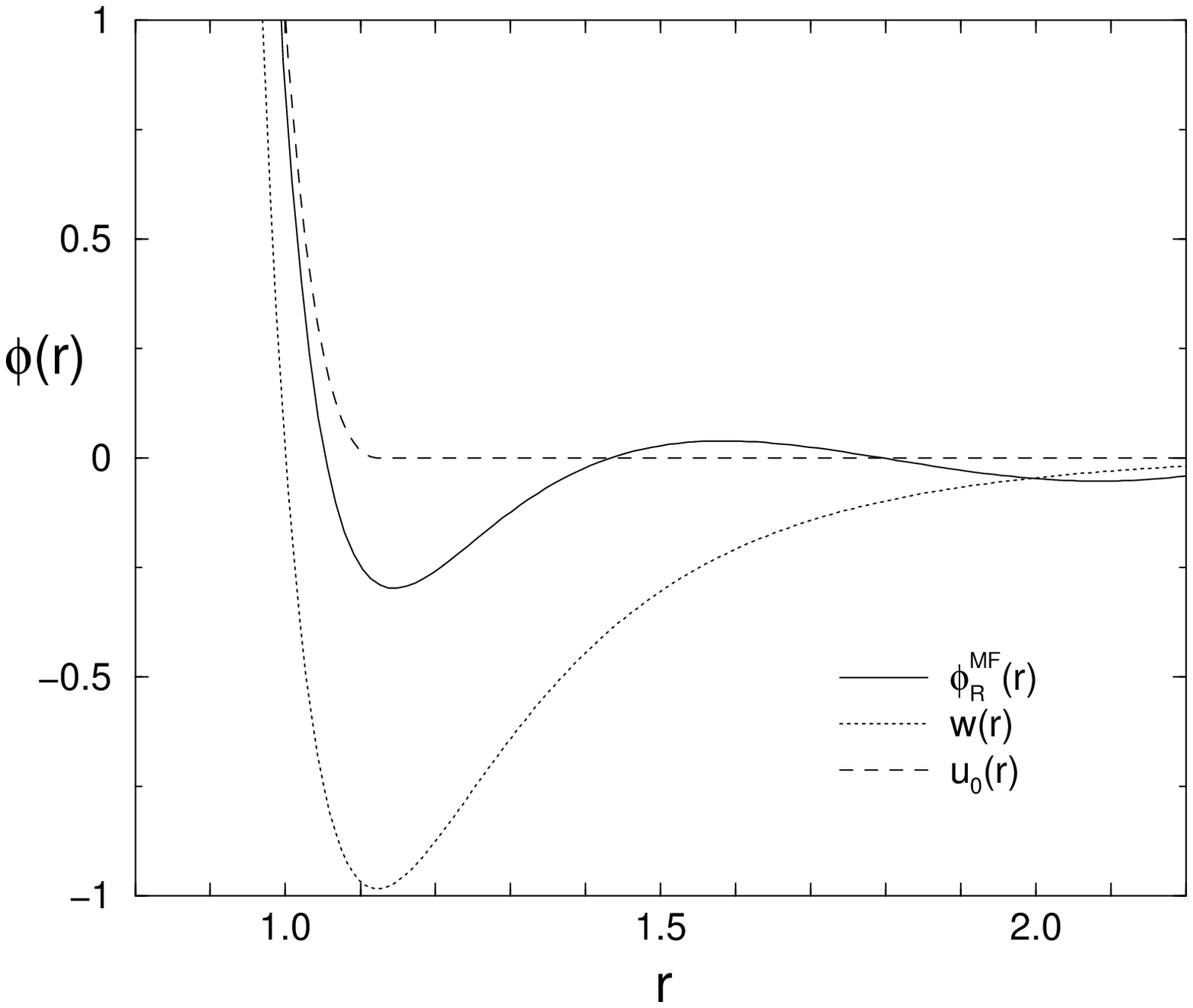}
\caption{The self-consistent ERF $\phi _{R}^{MF}(r)$ determined from the mean
field equation (\ref{mfint}) for $T=1.35$ and $\rho ^{B}=0.78$ compared to
the full LJ potential $w(r)$ and the repulsive force reference potential
$u_{0}(r).$}
\label{fig3}
\end{figure}
%%%%%%%%%%%%%
%%%%%%%%%%%%%

Figure (4) shows the ERF for a series of states along the $T=1.35$ isotherm.
The attractive force cancellation from further neighbors becomes
increasingly less effective at lower densities, and attractive interactions
produce much larger structural changes, as will be shown below.

%%%%%%%%%%%%%
%%%%%%%%%%%%%
\begin{figure}[t]
\epsfxsize=3.2in
\epsfbox{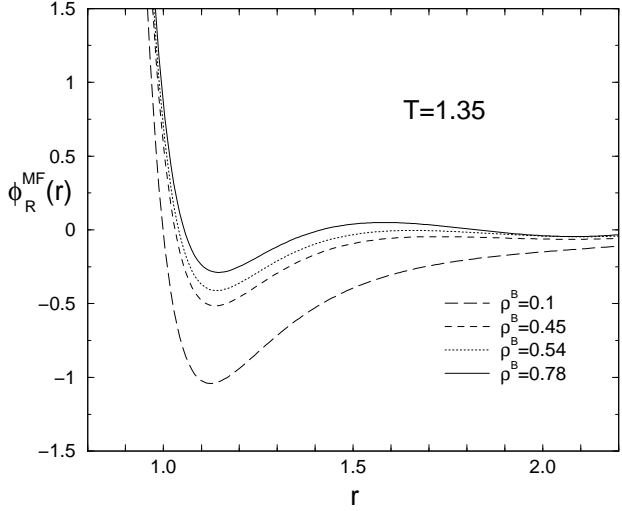}
\caption{The self-consistent ERF from Eq.~(\ref{mfint}) for the indicated
densities along the $T=1.35$ isotherm.}
\label{fig4}
\end{figure}
%%%%%%%%%%%%%
%%%%%%%%%%%%%

Again mean field theory can yield accurate results. This is illustrated in
Fig.~(5) for the low density state $\rho ^{B}=0.1$, and $T=1.35.$ This
figure shows simulation results for the full LJ density $\rho ^{B}g(r),$ the
WCA repulsive fluid density $\rho ^{B}g_{0}(r),$ and the inhomogeneous
reference density $\rho _{0}(r;[\phi _{R}^{MF}])$. Also shown is $\rho
_{0}(r;[\phi _{R}^{IMF}]),$ with the ERF calculated from Eq.~(\ref{IMF}) in
Appendix A (using $I_{2}(\rho )$ as the interpolation function), which does
a slightly better job at reproducing the second peak than does Eq.~(\ref
{mfint}). Theoretical values for correlation functions for all the states in
Fig.~(4) and comparison to results for the full LJ\ fluid will be discussed
in Sec.~\ref{results} below.

%%%%%%%%%%%%%
%%%%%%%%%%%%%
\begin{figure}[tbp]
\epsfxsize=3.2in
\epsfbox{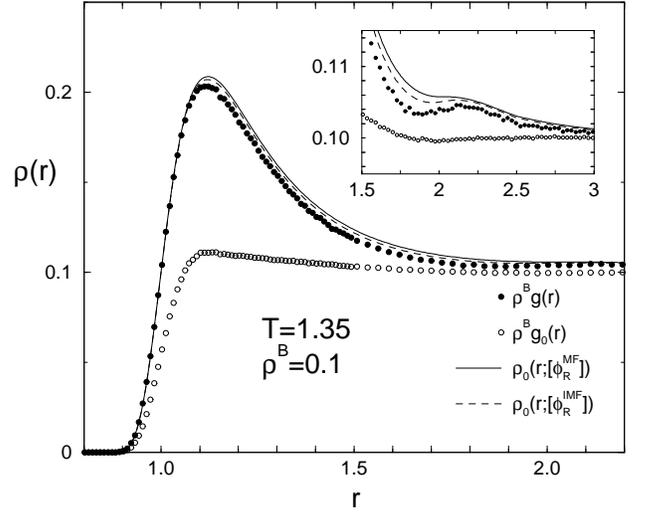}
\caption{Densities for $\rho ^{B}=0.10$ and $T=1.35$ for the LJ system, the
WCA reference system and the nonuniform reference system with the ERF
determined from Eq.~(\ref{mfint}) and from Eq.~(\ref{IMF}). The inset
focuses in the region around the second peak.}
\label{fig5}
\end{figure}
%%%%%%%%%%%%%
%%%%%%%%%%%%%

As the density tends to zero, the ERF reduces to the bare potential $w(r)$
as correctly predicted by Eq.~(\ref{mfint}). The increased ``screening'' of
attractive forces as the density is increased was first demonstrated using
diagrammatic resummation techniques in the derivation of the ORPA and EXP
integral equations \cite{ac72}. Mean field theory provides a very simple and
physically suggestive way of understanding these results.

\section{Two step method}

\label{TwoStepSec} We now discuss theoretical methods \cite{wkv,kw} for
determining the density $\rho _{0}({\bf r};{\bf [}\phi _{R}])\equiv $ $\rho
_{R}({\bf r})$ produced by a given ERF $\phi _{R}$. We use a generalization
of the two step method first introduced in Ref. \cite{wkv}. Initially we
treat the LJ repulsive potential $u_{0}$ as a hard core interaction with
diameter $d$, but then use the standard ``blip function'' expansion \cite
{hansenmac} to correct for its finite softness in our final numerical
results. Let us concentrate on the mean field equation (\ref{mfint}) for the
ERF. Recall that $\phi _{LJ}(r)\equiv w(r)=u_{0}(r)+u_{1}(r)$.

\subsection{Separation of ERF}

The two step method introduces a similar division of the full ERF, 
\begin{equation}
\phi _{R}(r)\equiv \phi _{R0}(r)+\phi _{R1}(r),  \label{phi0phi1}
\end{equation}
and determines the density response to each part of the ERF in separate
steps. As this notation suggests, $\phi _{R0}$ is supposed to take account
of the harshly repulsive part of the ERF associated mainly with the
repulsive core of the fixed wall particle at the origin. The other component 
$\phi _{R1}$ is much more slowly varying and physically incorporates the
averaged effects of attractive interactions.

The mean field equation (\ref{mfint}) naturally separates into two such
parts by setting 
\begin{equation}
\phi _{R0}(r)\equiv u_{0}(r)  \label{B0}
\end{equation}
and 
\begin{equation}
\phi _{R1}(r)\equiv u_{1}(r)+\phi _{s}(r),  \label{B1}
\end{equation}
with $\phi _{s}$ given by Eq.~(\ref{phis}). We call this the {\em basic }
separation and will use it in most of what follows.

However, other choices can be made. As discussed in Appendix B, there exists
considerable freedom to vary $\phi _{R1}(r)$ inside the harshly repulsive
core region where $u_{0}(r)$ is very large, without affecting the final
result when both parts of the ERF are taken into account. This flexibility
can be used to increase the accuracy of approximations introduced there that
require a slowly varying density response.

\subsection{ First step}

\label{firststep} The key idea in the two step method is to compute in an
initial step the density response to the {\em slowly varying part} $\phi
_{R1}(r)$ of the ERF {\em alone}. Physically, this takes account of the
averaged effects of the attractive interactions modeled by $\phi _{R1}$ and
we can exploit the fact that the density response can be expected to be
reasonably slowly varying. The response to the remaining harshly repulsive
interaction $\phi _{R0}$ is then determined in a second step.

\subsubsection{Hydrostatic approximation}

Let us first consider the special case where $\phi _{R1}$ varies so slowly
that it is essentially constant over the range of a correlation length in
the bulk fluid. The associated density $\rho _{0}(r{\bf ;[}\phi _{R1}],\mu
_{0}^{B})$ is a functional of the external field $\phi _{R1}$ and a function
of the chemical potential $\mu _{0}^{B}$ and depends only on the difference
between these quantities. Thus for any fixed position $r_{1}$ we can define
a shifted chemical potential 
\begin{equation}
\mu _{0}^{r_{1}}\equiv \mu _{0}^{B}-\phi _{R1}(r_{1})  \label{hydromu}
\end{equation}
and shifted field 
\begin{equation}
\phi _{R1}^{r_{1}}(r)\equiv \phi _{R1}(r)-\phi _{R1}(r_{1}),
\label{hydrophi}
\end{equation}
whose parametric dependence on $r_{1}$ is denoted by a superscript, and we
have for all $r$ the exact relation 
\begin{equation}
\rho _{0}(r{\bf ;[}\phi _{R1}],\mu _{0}^{B})=\rho _{0}(r{\bf ;[}\phi
_{R1}^{r_{1}}],\mu _{0}^{r_{1}}).  \label{hydroexact}
\end{equation}
By construction the shifted field $\phi _{R1}^{r_{1}}(r)$ vanishes at $r=$ $%
r_{1}\ $and it remains very small for $r$ near $r_{1}$ when $\phi _{R1}$ is
very slowly varying. Thus to determine the density at $r_{1}$ we can
approximate the R.H.S. of Eq.~(\ref{hydroexact}) by $\rho _{0}(r_{1}{\bf ;[}%
0],\mu _{0}^{r_{1}})\equiv \rho _{0}(\mu _{0}^{r_{1}})$, the density of the 
{\em uniform} fluid (in zero field) at the shifted chemical potential $\mu
_{0}^{r_{1}}.$ We arrive at the{\em \ hydrostatic} {\em approximation} \cite
{frishleb,localdensity} for the density arising from a very slowly varying
field $\phi _{R1}$: 
\begin{equation}
\rho _{0}(r_{1}{\bf ;[}\phi _{R1}],\mu _{0}^{B})\;\approx \;\rho _{0}(\mu
_{0}^{r_{1}}).  \label{hydrorho}
\end{equation}
We refer to $\rho _{0}(\mu _{0}^{r_{1}})\equiv \rho _{0}^{r_{1}}$ as the 
{\em hydrostatic density} at $r_{1}$; from Eqs.~(\ref{hydromu}) and (\ref
{hydrorho}) it depends only on the {\em local} value of the field $\phi _{R1}
$ at $r_{1}.$ The hydrostatic approximation is exact for sufficiently slowly
varying $\phi _{R1}$ and has been used in more approximate applications of
these ideas to hydrophobic interactions in water \cite{lcw}. However in the
present application $\phi _{R1}$ varies rapidly enough that for quantitative
accuracy we must use more accurate methods to determine the full nonlocal
response.

We now show that the generalized linear response method introduced in Ref. 
\cite{kw} provides a simple and accurate way to determine both the density $%
\rho _{R1}$ induced in the first step as well as the response to $\phi _{R0}$
taken into account in the second step. In Appendix B we discuss an alternate
but somewhat more complicated approach suggested in Ref. \cite{wkv}, which
requires that $\rho _{R1}$ is sufficiently slowly varying that gradient type
expansions give accurate results.

We start from the exact linear response equation that relates {\em small}
changes in the potential and density for a system with external potential $%
\phi $, chemical potential $\mu $, and associated density $\rho _{0}(r;[\phi
],\mu )\equiv \rho _{\phi }(r)$ \cite{hansenmac,wkv}: 
\begin{equation}
-\beta \delta \phi ({\bf r}_{1})=\int \!d{\bf r}_{2}\,\chi _{0}^{-1}({\bf r}
_{1},{\bf r}_{2};{\bf [}\rho _{\phi }])\delta \rho _{\phi }(r_{2})
\label{linresponse}
\end{equation}
through the linear response function $\chi _{0}^{-1}({\bf r}_{1},{\bf r}_{2};%
{\bf [}\rho _{\phi }])\equiv \delta ({\bf r}_{1}\!-\!{\bf r}_{2})/\rho
_{\phi }({\bf r}_{1})\!-\!c_{0}({\bf r}_{1},{\bf r}_{2};{\bf [}\rho _{\phi
}])$. Here $\!c_{0}$ is the direct correlation function of the system with
density $\rho _{\phi }(r).$

\subsubsection{Linear Response of Hydrostatic Fluid}

\label{linhydrosec} The simple hydrostatic method discussed above
approximates $\rho _{R1}(r_{1})$ at each $r_{1}$ by the density $\rho
_{0}^{r_{1}}$ of the uniform hydrostatic fluid with chemical potential $\mu
_{0}^{r_{1}},$ and thus ignores the {\em nonlocal} effects of the shifted
field $\phi _{R1}^{r_{1}}$ on the density at $r_{1}.$ To get a more accurate
approximation, we can use Eq.~(\ref{linresponse}) to take into account the 
{\em linear response }of the density of the uniform hydrostatic fluid to the
shifted field $\phi _{R1}^{r_{1}}.$ Thus we set $\rho _{\phi }=\rho
_{0}^{r_{1}}$ and take $\delta \phi =\phi _{R1}^{r_{1}}$ in (\ref
{linresponse}). This idea was first suggested in Ref. \cite{kw} and was
shown to give accurate results in a number of different applications. While
a more formal derivation can be given \cite{kw}, here we focus on physical
considerations.

Since $\phi _{R1}^{r_{1}}(r)$ is zero at $r_{1},$ the left side of Eq.~(\ref
{linresponse}) vanishes by construction. We would expect the linear response
relation between an external field and induced density to be most accurate
where the field is small --- in particular where the field vanishes --- and
at each $r_{1}$ we will use the appropriate shifted (hydrostatic) chemical
potential and shifted field so that this optimal condition continues to hold
locally. This shift is crucial for the accuracy of this method and is its
main new feature over previous approaches. Moreover, it has been shown that
even large density fluctuations in a (field free) hard sphere fluid can be
accurately described using the {\em same} Gaussian probability distribution
that controls small fluctuations \cite{crooks} and that yields the basic
linear response relation (\ref{linresponse}) for a uniform system.

This suggests that we can accurately determine the desired $\rho _{R1}(r_{1})
$ by using the linear response function $\chi _{0}^{-1}({\bf r}_{12};\rho
_{0}^{r_{1}})$ of the uniform hydrostatic fluid in Eq.~(\ref{linresponse})
even when the field $\phi _{R1}^{r_{1}}$ produces significant density
changes. Assuming a linear density response, we replace $\delta \rho _{\phi
}(r)$ in (\ref{linresponse}) by the full density change $\rho _{R1}(r)-\rho
_{0}^{r_{1}}$, thus yielding our final result: 
\begin{equation}
\lbrack \rho _{R1}(r_{1})-\rho _{0}^{r_{1}}]/\!\rho _{0}^{r_{1}}=\int \!d%
{\bf r}_{2}\,c_{0}(r_{12};\rho _{0}^{r_{1}})[\rho _{R1}(r_{2})-\rho
_{0}^{r_{1}}]\;.  \label{linhydro}
\end{equation}
Here $c_{0}(r_{12};\rho _{0}^{r_{1}})$ is the direct correlation of the
uniform reference fluid at the hydrostatic density $\rho _{0}^{r_{1}}.$ Note
that the external field appears only implicitly in Eq.~(\ref{linhydro})
through its local effect on the hydrostatic density $\rho _{0}^{r_{1}}$.

Equation (\ref{linhydro}) is a {\em linear} integral equation relating the
density $\rho _{R1}(r_{1})$ at a given $r_{1}$ on the left side to an
integral involving the density $\rho _{R1}(r_{2})$ at all other points and a 
{\em uniform fluid} kernel $c_{0}(r_{12};\rho _{0}^{r_{1}})$ that depends
implicitly on $r_{1}$ through $\rho _{0}^{r_{1}}.$ This new feature presents
no technical difficulties in determining a numerical solution and Eq.~(\ref
{linhydro}) can be solved by any number of standard methods. We found that
Picard iteration works very well. See Appendix C for details.

\subsection{Second step}

\label{secondstep} We now determine in a second step the response $\Delta %
\rho _{R}(r)\equiv \rho _{0}(r{\bf ;[}\phi _{R}])-\rho _{0}(r{\bf ;[}\phi
_{R1}])$ of the relatively slowly varying density field $\rho _{0}(r{\bf ;[}%
\phi _{R1}])\equiv \rho _{R1}(r)$ to the remaining harshly repulsive
component $\phi _{R0}$ of the ERF, which we approximate initially as a hard
core of range $d$. We take $\rho _{\phi }=\rho _{R1}$ in Eq.~(\ref
{linresponse}) and again assume a linear density response in the ``out''
region $r_{1}>d$ where the perturbing potential $\phi _{R0}(r_{1})$
vanishes. This is consistent with the simulation results \cite{crooks}
showing that the Gaussian probability distribution gave a good description
even of the formation of voids in uniform fluids.

This linear response assumption gives the approximate equation, valid for $%
r_{1}>d$: 
\begin{equation}
0=\int \!d{\bf r}_{2}\,\chi _{0}^{-1}({\bf r}_{1},{\bf r}_{2};{\bf [}\rho
_{R1}])\Delta \rho _{R}(r_{2}),  \label{linresp2}
\end{equation}
where we impose the exact condition $\rho _{0}(r_{2}{\bf ;[}\phi _{R}])=0$
from the hard core interaction for $r_{2}<d$ in the integration over $r_{2.}$
Again we approximate $\chi _{0}^{-1}({\bf r}_{1},{\bf r}_{2};{\bf [}\rho
_{R1}])$ by the response function of an appropriately chosen uniform system.
As in Sec.~\ref{linhydrosec} and as shown in Ref. \cite{kw}, we find that
the use of the hydrostatic fluid with density $\rho _{0}^{r_{1}}$ gives
accurate results even when $\rho _{R1}$ varies rather rapidly. Some
alternate but less generally useful choices are discussed in Appendix B.
Thus we arrive at the basic equation for the second step of our theory,
valid for $r_{1}>d:$%
\begin{equation}
\Delta \rho _{R}(r_{1})/\!\rho _{0}^{r_{1}}=\int \!d{\bf r}%
_{2}\,c_{0}(r_{12};\rho _{0}^{r_{1}})\Delta \rho _{R}(r_{2}).  \label{Wallpy}
\end{equation}

Equation (\ref{Wallpy}) is a {\em linear} equation for $\Delta \rho
_{R}(r_{1}),$ which we can directly solve by iteration or other means. When $%
\rho _{R1}(r{\bf )}=\rho ^{B},$ and $c_{0}$ is assumed to vanish for $r>d$,
Eq.~(\ref{Wallpy}) reduces to the standard PY equation \cite{PY} for the
uniform hard sphere fluid. This has an analytic solution \cite
{frishleb,hansenmac} and is known to give very good results overall, with
small errors in the height of the first peak at very high densities. If
still more accuracy is required, we can use modified GMSA type equations 
\cite{gmsawais} related to the PY equation to describe $c_{0}$, as discussed
in Appendix D. Again we can solve Eq.~(\ref{Wallpy}) by iteration.

This constitutes the second step of our method. The net result of this two
step process is the desired $\rho _{0}(r{\bf ;[}\phi _{R}])$ arising from a
given $\phi _{R}.$ This can be substituted into Eq.~(\ref{mfint}), which can
then be iterated to determine the final self-consistent $\phi _{R}^{MF}$ and 
$\rho _{0}(r{\bf ;[}\phi _{R}^{MF}])$. See Appendix C for further details of
the calculations.

\section{Results}

%%%%%%%%%%%%%
%%%%%%%%%%%%%
\begin{figure}[t]
\epsfxsize=3.2in
\epsfbox{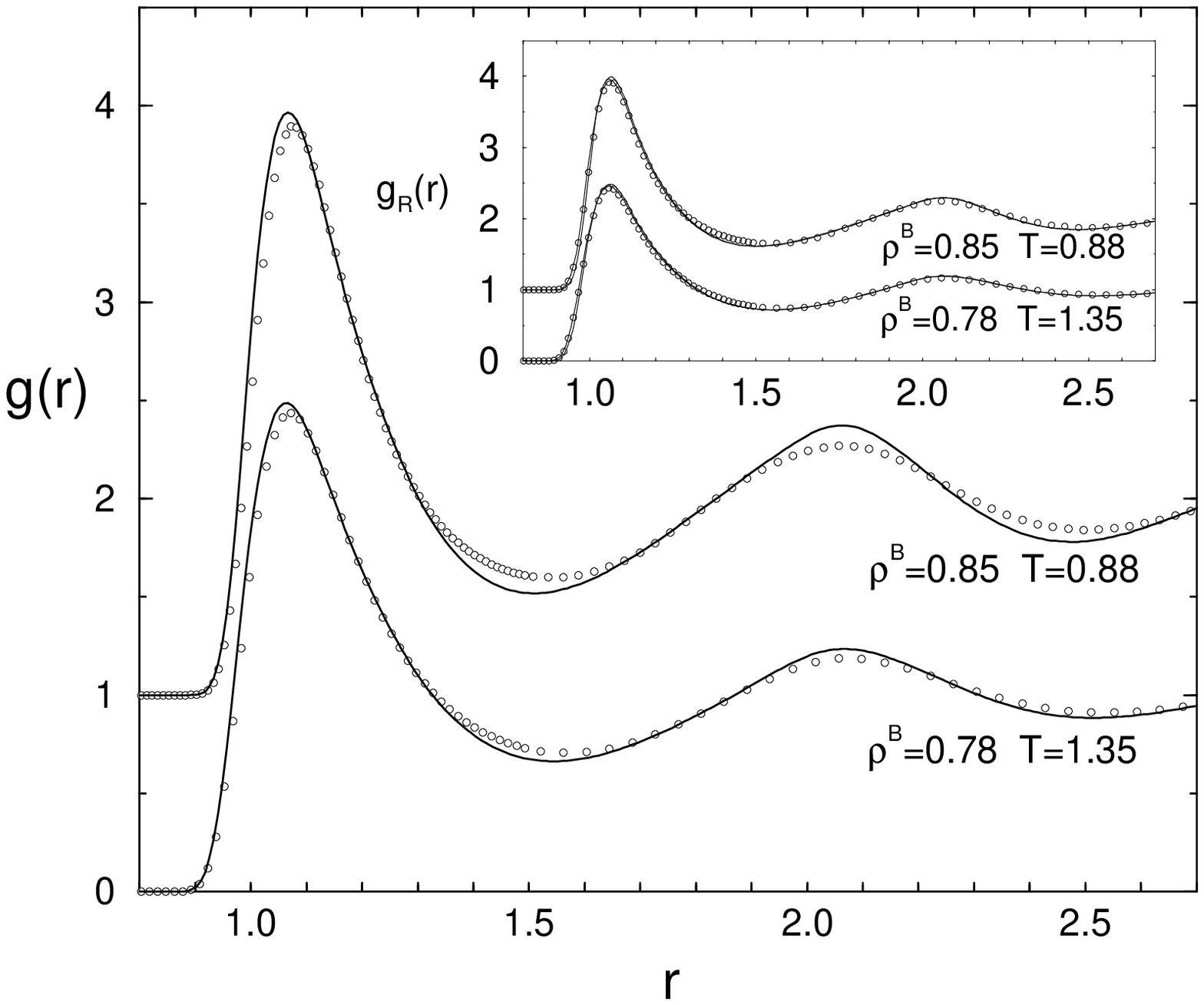}
\caption{ \mbox{Radial distribution functions $g_{0}(r_{2};{\bf [}\phi _{R}^{MF}])$}
$=\rho _{0}(r_{2};{\bf [}\phi _{R}^{MF}])/\rho ^{B}$
for two high density states (solid lines with the origin shifted for clarity) as given by
the two step method using Eq.~(\ref{mfint}) for the ERF along with Eqs.~(\ref
{linhydro}) and (\ref{Wallpy}) compared to the results of MD simulations
(open circles) of the {\em LJ fluid}. In the inset the two step results are
compared directly to simulations of the {\em reference fluid} in the
self-consistent potential $\phi _{R}^{MF}.$ The latter tests the accuracy of
the two-step method for reference fluid correlations induced by the given $%
\phi _{R}^{MF}$, while the former tests the accuracy of the mean field
equation for determining $\phi _{R}^{MF}.$}
\label{fig6}
\end{figure}
%%%%%%%%%%%%%
%%%%%%%%%%%%%

\label{results} We now give a detailed comparison of the radial distribution
functions $g_{0}(r_{2};{\bf [}\phi _{R}^{MF}])=\rho _{0}(r_{2};{\bf [}\phi
_{R}^{MF}])/\rho ^{B}$ given by the two step method to the results of MD
simulations of the full LJ fluid. In Fig.~(6) we consider two high density
states with $\rho ^{B}=0.78$ and $T=1.35$ and $\rho ^{B}=0.85$ and $T=0.88$.
At these high densities small errors can be seen in the linear response
treatment (equivalent to the hard core PY equation) of even the uniform hard
sphere reference fluid. For greater accuracy therefore we used an improved
GMSA description as briefly described in Appendix D. We find by direct
comparison with simulations of the reference fluid in presence of the
self-consistent ERF that the two step method indeed gives a very accurate
description of the nonuniform reference fluid density, as illustrated in the
inset to Fig.~(6).

The main remaining errors in describing the full LJ fluid are thus
associated with the mean field approximation for the ERF. As already
indicated in Fig.~(2), this properly describes the basic shift in the first
peak when compared to the WCA uniform reference fluid, but it also produces
slight overshoots in the peaks and minima for these high density states, as
can be seen in Fig.~(6).

Figure (7) gives results for states at $T=1.35$ and a series of lower
densities: $\rho ^{B}=0.54$, $\rho ^{B}=0.45,$ and $\rho ^{B}=0.10.$ Here
the linear response treatment of the reference system is sufficiently
accurate and the attractive interactions produce large changes in the
correlation functions. The results seem quite satisfactory, and are
comparable to those given by the best standard integral equation methods 
\cite{lado73,hmsa}. Results from the alternate methods discussed in Appendix
B are equally good and essentially indistinguishable on the scale of the
graph.

%%%%%%%%%%%%%
%%%%%%%%%%%%%
\begin{figure}[t]
\epsfxsize=3.2in
\epsfbox{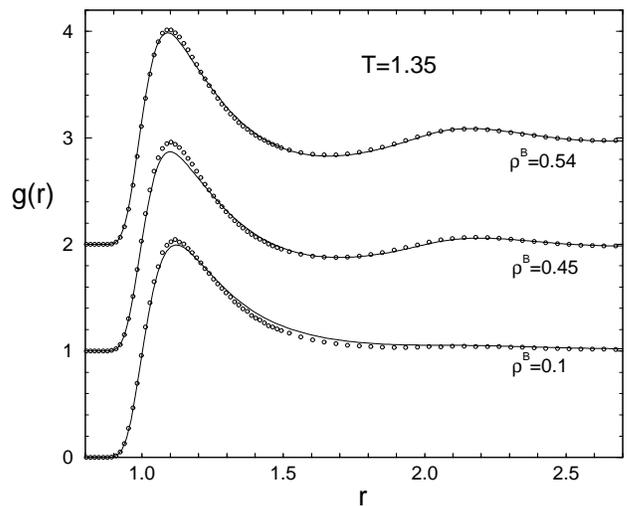}
\caption{Radial distribution functions for the lower density states indicated
compared to the results of MD simulations (open circles) of the LJ fluid.
The symbols have the same meaning as in Fig.~6.}
\label{fig7}
\end{figure}
%%%%%%%%%%%%%
%%%%%%%%%%%%%

\section{Final Remarks}

\label{Finalremarks} These results thus give us additional confidence in the
utility of our general approach. While we certainly do not advocate
replacing standard and successful integral equation methods for the specific
problem of the structure of uniform simple fluids, these ideas do suggest
new ways of thinking about some basic issues. We can view the simple mean
field approximation for the attractive interactions along with the
generalized linear response treatment of correlations in the reference fluid
as providing reasonably accurate and computationally practical first
approximations for correlations induced by attractive and repulsive
interactions. For qualitative and often quantitative work they have proved
useful in a variety of different applications, including cases such as
drying near walls \cite{wkv} where attractive forces induce large structural
changes and standard integral equation methods fail.

For quantitative accuracy, improvements in both approximations may be called
for in some cases. Incorporation of GMSA type corrections \cite{gmsawais}
for reference fluid correlations is straightforward, as discussed in
Appendix D, and alternate and probably more accurate treatments of the
effects of soft cores can be used if needed \cite{lit:KatVolWee}. Some
corrections to the simplest mean field equation for the ERF, as discussed in
Appendix A or in Ref. \cite{wvk}, can also be introduced. However, there are
some fundamental errors arising from the use of any mean field approximation
for the attractive interactions that cannot be easily avoided. The inherent
limitations of mean field theory in treating long wavelength correlations
such as those seen at the critical point or arising from capillary waves at
the liquid-vapor interface are well known. Fortunately in many applications
of interest such correlations do not play an important role, or their
effects can be taken into account separately. In such cases the ideas
discussed here provide a unified and physically suggestive perspective
capable of giving a good qualitative and often a quantitative description of
the structure of both uniform and nonuniform fluids.

\section{Acknowledgments}

This work was supported by the National Science Foundation, Grant No.
CHE9528915. While at University of Maryland, KVL also received support from
the Deutsche Forschungsgemeinschaft.

\appendix

\section{Interpolated mean field equation}

Equation (\ref{mfint}) is {\em exact} as the density $\rho $ tends to zero,
where $\phi _{R}^{MF}$ reduces to the bare field $\phi _{LJ}$. However, the
next order term [of $O(\rho )$] in a density expansion is incorrect. This
can most easily be seen by comparing the known density expansions \cite
{hansenmac} for the LJ system's $\rho (r_{1};{\bf [}\phi _{LJ}])$ with a LJ
particle fixed at the origin, and the reference system's $\rho _{0}(r_{1};%
{\bf [}\phi _{R}])$ with the special wall particle with pair interaction $%
w_{R}(r_{01})=\phi _{R}(r_{1})$ fixed at the origin.

We examined an empirical modification of Eq.~(\ref{mfint}) that at low
density gives the next term of $O(\rho )$ exactly but then quickly goes over
to Eq.~(\ref{mfint}) at higher density: 
\begin{eqnarray}
\beta \phi _{R}^{IMF}(r_{1}) &=&\beta \phi _{LJ}(r_{1})-\int d{\bf r}%
_{2}\,\{[\rho _{0}(r_{2};{\bf [}\phi _{R}^{IMF}])-\rho ^{B}]  \nonumber \\
&&\times [1+I(\rho )f_{0}(r_{12})]F_{1}(r_{12};\rho ^{B})\}.  \label{IMF}
\end{eqnarray}
Here $I(\rho )$ is an {\em interpolation function} that tends to unity at
low density and to zero at high density, $f_{0}(r)\equiv \exp [-\beta
u_{0}(r)]-1$, and 
\begin{equation}
F_{1}(r;\rho )\equiv [\exp \{-\beta u_{1}(r)I(\rho )\}-1]/I(\rho ).
\label{F1}
\end{equation}

Possible choices for $I(\rho )$ include $I_{1}(\rho )=(\partial \rho /%
\partial \beta p)_{0}$ $\equiv S_{0,\rho }$ \cite{davis}, proportional to
the reference fluid isothermal compressibility, and $I_{2}(\rho )=$ $%
S_{0,\rho }^{2}$, as suggested by a crude argument \cite{crudeIMF} based on
perturbing the hard sphere Ornstein-Zernike equation by a very weak and
slowly varying potential. At the lowest density studied, $\rho =0.1,$ Eq.~(%
\ref{IMF}) with $I_{2}(\rho )$ gave a slightly better description of the
weak second peak than does the simple mean field equation (\ref{mfint}). See
Fig.~(5).

\section{Alternate equations for slowly varying $\rho _{R1}.$}

\label{slowlyvarying} In Sec.~\ref{TwoStepSec} we exploited the Gaussian
nature of fluctuations in the uniform reference system in carrying out both
steps of the two step method. While this is a good approximation for LJ
reference system, it may not always hold true. In our initial work in Ref. 
\cite{wkv}\ we proposed a different, and very general way of carrying out
the first step, which however requires that $\rho _{R1}$ varies sufficiently
slowly that gradient type expansions give good results.

We started from an exact equation \cite{derivelmb} first derived by Lovett,
Mou and Buff (LMB) \cite{lmb}: 
\begin{eqnarray}
\nabla _{1}\rho _{R1}(r_{1}) &/&\rho _{R1}(r_{1})=-\beta \nabla _{1}\phi
_{R1}(r_{1})  \nonumber \\
&+&\int d{\bf r}_{2}c_{0}({\bf r}_{1},{\bf r}_{2};{\bf [}\rho _{R1}])\nabla
_{2}\rho _{R1}(r_{2}).\,  \label{LMBequation}
\end{eqnarray}
The $c_{0}({\bf r}_{1},{\bf r}_{2};{\bf [}\rho _{R1}])$ for a general
nonuniform $\rho _{R1}$ is difficult to determine, so Eq.~(\ref{LMBequation}%
) is generally not very useful for practical calculations. However, if $\rho
_{R1}$ is relatively {\em slowly varying}, then we can accurately
approximate $c_{0}({\bf r}_{1},{\bf r}_{2};{\bf [}\rho _{R1}])$ under the
integral in Eq.~(\ref{LMBequation}) by the {\em uniform fluid} function $%
c_{0}(r_{12};\bar{\rho}_{12})$ \cite{cisbetter}, where $\bar{\rho}_{12}$ is
some average density associated with the two points. Then Eq.~(\ref
{LMBequation}) can be solved to determine $\rho _{R1}.$

A natural choice for $\bar{\rho}_{12}$ suggested by a gradient expansion 
\cite{frishleb} is $\bar{\rho}_{12}=[\rho _{R1}(r_{1})+\rho _{R1}(r_{2})]/2$
. This gives very good results when $\rho _{R1}$ is reasonably smooth. This
is the only approximation we make and we can check its accuracy by seeing if
similar results arise from other approximations such as $\bar{\rho}%
_{12}=\rho _{R1}(r_{1})$ or $\bar{\rho}_{12}=\rho _{R1}(r_{2}).$ Starting
with a given $\phi _{R1}$, we can then solve Eq.~(\ref{LMBequation}) for the
associated $\rho _{R1}$ by iteration, making use of the analytic and
accurate Percus-Yevick (PY) expressions for the direct correlation function
of the uniform hard sphere fluid \cite{frishleb,PY}. If more accuracy is
required we can use GMSA type equations related to the PY equation \cite
{gmsawais} to describe $c_{0}$. See Appendix D.

In our previous study of the LJ fluid near a hard wall \cite{wkv}, we used
the {\em basic} separation of $\phi _{R}$ in Eqs.~(\ref{B0}) and (\ref{B1}),
and found that it indeed produced a slowly varying density response. As
expected Eq.~(\ref{LMBequation}) then gave very accurate results. However,
in the present application, the size of the excluded volume region of the
fixed particle (of order $\sigma $ of the LJ potential) is also the same
order as the range of the attractive interactions as well as the average
spacing between particles at high density. If the basic separation is used,
this ``resonance'' produces a $\phi _{R1}^{B}$ at very high density with
small but noticeable oscillations of period $2\pi /\sigma $ outside the core
and a pronounced minimum inside the core at $r_{1}=0.$ The associated
density response $\rho _{0}(r_{1};{\bf [}\phi _{R1}^{B}])$ will have a
pronounced maximum at $r_{1}=0$ and oscillations outside the core, which
will cause errors in the local expansion method used in Eq.~(\ref
{LMBequation}) and in Eq.~(\ref{oldwallpy}) below.

Fortunately we can use the flexibility in the choice of the field $\phi
_{R1}(r)$ in the core region to produce a much smoother density response.
More precisely, we can define an {\em extended }separation of $\phi _{R}(r)$
in Eq.~(\ref{phi0phi1}) by: 
\begin{equation}
\phi _{R0}^{E}(r)\equiv u_{0}(r)-\phi _{0}^{E}(r)  \label{E0}
\end{equation}
and 
\begin{equation}
\phi _{R1}^{E}(r)\equiv u_{1}(r)+\phi _{s}(r)+\phi _{0}^{E}(r),  \label{E1}
\end{equation}
where $\phi _{0}^{E}(r)$ is an essentially arbitrary smooth function that is
nonzero only in the repulsive core region but with $\phi
_{0}^{E}(r)<<u_{0}(r)$, so that $\phi _{R0}^{E}$ remains a harshly repulsive
(essentially hard core) interaction. This{\em \ }separation still divides
the ERF $\phi _{R}(r)$ into two parts with the physical meaning discussed in
Sec.~\ref{TwoStepSec}, but provides some additional flexibility in the
choice of $\phi _{R1}(r)$ in the core region that can be used to produce a
smoother density response in the first step. An exact treatment of the
response to both components of $\phi _{R}(r)$ would of course be independent
of how the potential was separated.

We found best results by requiring that density response to $\phi _{R1}^{E}$
inside the (hard) core region be {\em constant }and continuous across the
core. In a sense this is the smoothest possible choice, at least in the
vicinity of the core region. This choice can easily be implemented
numerically during the iterative solution of Eq.~(\ref{LMBequation}) by
simply setting $\nabla _{1}\rho _{R1}(r_{1})$ to be zero for all $r_{1}$
inside the core on each iteration and solving for the associated $\nabla
_{1}\phi _{R1}^{E}(r_{1})$. At convergence, the self-consistent $\rho
_{R1}^{E}(r)$ is constant inside the core and smoother outside the core than
that produced by the basic separation.

Using the same extended separation, we have verified by comparison with the
hydrostatic linear response method and with direct simulations that Eq.~(\ref
{LMBequation}) now gives accurate results for all the states tested here.
Thus it offers an alternative (though numerically slightly more complicated)
way of carrying out the first step.

In Ref. \cite{wkv} we also carried out the second step in a slightly
different way, effectively combining a local expansion of $\rho _{R1}$ with
a linear response treatment of the density induced by $\phi _{R0}$. In
particular, in Eq.~(\ref{linresp2}) we treated the $\delta $-function part
of $\chi _{0}^{-1}({\bf r}_{1},{\bf r}_{2};{\bf [}\rho _{R1}])$ exactly and
approximated the $c_{0}$ part by the uniform fluid function at the
intermediate density $\bar{\rho}_{12}=[\rho _{R1}(r_{1})+\rho
_{R1}(r_{2})]/2.$ The accuracy of this approximation can again be checked by
comparing with other choices such as $\bar{\rho}_{12}=\rho _{R1}(r_{1}).$
This yields the alternate equation for the second step: 
\begin{equation}
\Delta \rho _{R}(r_{1})/\!\rho _{R1}(r_{1})=\int \!d{\bf r}%
_{2}\,c_{0}(r_{12};\bar{\rho}_{12})\Delta \rho _{R}(r_{2}).
\label{oldwallpy}
\end{equation}

When $\rho _{R1}(r_{1})$ is slowly varying, as was the case in all the
examples studied in the previous work, Eq.~(\ref{oldwallpy}) gives accurate
results, essentially identical to those of Eq.~(\ref{Wallpy}). This is also
true for most of the states studied in the present application, provided
that the appropriate extended separation is used. In such cases Eqs.~(\ref
{LMBequation}) and (\ref{oldwallpy})\ can be used as alternate ways of
implementing the two step method, and for the states shown in Fig.~(7) they
give results on the scale of the graph essentially identical to those shown.
However for the high density states $\rho ^{B}=0.78$ and $T=1.35$ and $\rho
^{B}=0.85$ and $T=0.88$ in Fig.~(6) the results using Eq.~(\ref{oldwallpy})
vary significantly when different choices for $\bar{\rho}_{12}$ are made.
This indicates that for these states $\rho _{R1}^{E}$ is too rapidly varying
for Eq.~(\ref{oldwallpy}) to be trusted. Since Eq.~(\ref{Wallpy}) gives
accurate results even for these high density states, and makes fewer
assumptions about the smooth behavior of $\rho _{R1},$ it is the preferred
way to carry out the second step \cite{basicokforlinres}.

\section{Details of The Numerical Calculations}

We give here some details of the numerical solution of the basic equations (%
\ref{mfint}), (\ref{linhydro}), and (\ref{Wallpy}). Eqs.~(\ref{LMBequation})
and (\ref{oldwallpy}) could be used as alternates in the first and second
step respectively except at the highest densities. We exploit the spherical
symmetry of the density and the ERF about the center of the fixed wall
particle, which we take as the origin of a spherical coordinate system.
Since all these equations are used iteratively, we need an efficient and
accurate method to calculate three dimensional integrals over ${\bf r}_{2}$
of the general form 
\begin{equation}
I(r_{1})=\int d{\bf r}_{2}\,k(r_{2})K(r_{12};\overline{\rho }),  \label{I3}
\end{equation}
where due to the spherical symmetry $k(r_{2})$ depends only on $r_{2}\equiv |%
{\bf r}_{2}|$, and $K(r_{12};\overline{\rho })$ is a function only of $%
r_{12}\equiv |{\bf r}_{1}-{\bf r}_{2}|$ and $r_{1}$ and $r_{2}$ through our
choice of the effective density $\overline{\rho }$ which equals either the
hydrostatic density $\overline{\rho }=\rho _{0}^{r_{1}}$ or the average
density $\overline{\rho }=\overline{\rho }_{12}\equiv [\rho
_{R1}(r_{1})+\rho _{R1}(r_{2})]/2$.

These properties make it advantageous for us to use {\em bipolar coordinates}
\cite{bipolar} with the substitution $%
y^{2}=r_{12}^{2}=r_{1}^{2}+r_{2}^{2}-2r_{1}r_{2}\cos \theta $ and reduce the
three dimensional integration to two: 
\begin{equation}
I(r_{1})=\frac{2\pi }{r_{1}}\,\int\limits_{0}^{\infty }dr_{2}\,r_{2}%
\,k(r_{2})\,\int\limits_{|r_{1}-r_{2}|}^{r_{1}+r_{2}}dy\,y\,K(y;\overline{%
\rho }).  \label{3to2dint}
\end{equation}
This transformation is particularly useful if the dependence of $K(y;%
\overline{\rho })$ on $y$ is known analytically, since then we can
explicitly carry out the $y$ integration, and Eq.~(\ref{3to2dint}) further
reduces to a one dimensional integral. All relevant equations have $K$'s
that satisfy this condition, thus permitting very efficient numerical
computations.

In particular, Eq.~(\ref{mfint}) becomes: 
\begin{eqnarray}
\phi _{R}^{MF}(r_{1})-\phi _{LJ}(r_{1}) &=&\frac{2\pi }{r_{1}}\int_{0}^{%
\infty }dr_{2}\,r_{2}\,[\rho _{0}(r_{2};{\bf [}\phi _{R}^{MF}])-\rho ^{B}] 
\nonumber \\
&&\times \int\limits_{|r_{1}-r_{2}|}^{r_{1}+r_{2}}dy\,y\,u_{1}(y),
\label{mfintT}
\end{eqnarray}
while Eq.~(\ref{Wallpy}) is transformed to: 
\begin{eqnarray}
\Delta \rho _{R}(r_{1})/\!\rho _{0}^{r_{1}} &=&\frac{2\pi }{r_{1}}\int
\!dr_{2}\,r_{2}\Delta \rho _{R}(r_{2})  \nonumber \\
&&\times \int\limits_{|r_{1}-r_{2}|}^{r_{1}+r_{2}}dy\,y\,c_{0}(y;\rho
_{0}^{r_{1}}).  \label{wallpyT}
\end{eqnarray}
Equations (\ref{linhydro}) and (\ref{oldwallpy}) can be similarly rewritten.

The vector equation (\ref{LMBequation}) can be transformed into scalar form
by taking the scalar product with the unit vector ${\bf r}_{1}/r_{1}$ and
using the identity $2\,{\bf r_{1}}\cdot {\bf r_{2}}%
=r_{1}^{2}+r_{2}^{2}-r_{12}^{2}$. Thus we find: 
\begin{eqnarray}
\frac{d\ln \rho _{R1}(r_{1})}{dr_{1}} &=&-\frac{\beta d\phi _{R1}(r_{1})}{%
dr_{1}}+\frac{\pi }{r_{1}^{2}}\int \,dr_{2}\,\frac{d\rho _{R1}(r_{2})}{dr_{2}%
}  \nonumber \\
&&\times
\int\limits_{|r_{1}-r_{2}|}^{r_{1}+r_{2}}dy\,y%
\,(r_{1}^{2}+r_{2}^{2}-y^{2})c_{0}(y;\overline{\rho }_{12}).  \label{lmbeqnT}
\end{eqnarray}

In all these equations the integration over the variable $y$ can be
performed analytically, since we use the PY hard sphere direct correlation
function $c_{0}(y)$, which is a polynomial in $y$ \cite{hansenmac}, and the $%
u_{1}(y)$ of the Lennard-Jones potential, which is a sum of $y^{-6}$ and $%
y^{-12}$ terms \cite{cutoff}. Integrals involving an improved GMSA
approximation for $c_{0}$ can also be carried out analytically. The
resulting one dimensional integral equations can be solved by Picard
iteration, where to enforce convergence we use the usual mixing technique.

In solving these equations the reference potential $u_{0}$ is initially
taken to be a hard core potential with diameter $d$ given by the accurate
Verlet-Weiss expressions \cite{Verlet72}. As in the blip function method 
\cite{hansenmac}, the result for $\rho _{0}(r{\bf ;[}\phi _{R}])$ with $u_{0}
$ approximated by a hard core is linearly extrapolated into the core region
and multiplied by the Boltzmann factor of the true soft core potential $u_{0}
$ to give the final results for $\rho _{0}$ shown in Figs.~(6) and (7). The
errors introduced by this simplified treatment of soft cores are much
smaller than those arising from our use of the mean field approximation for
the ERF $\phi _{R}$.

\section{Generalized Mean Spherical Treatment of Reference System}

The description of the reference system presented here relies on accuracy of
the uniform hard sphere fluid direct correlation function $c_{0}(r)$. The
generalized linear response treatment of Eq.~(\ref{Wallpy}) with $\rho
_{0}^{r_{1}}=\rho ^{B}$ and $c_{0}$ vanishing outside the core is equivalent
to the PY approximation, and is surprisingly accurate at intermediate and
low densities. However, at high density it has noticeable errors, especially
in the region of the first peak near contact, and for quantitative results
should be corrected.

Since the uniform fluid direct correlation function is just an input in our
approach we can use other, more accurate approximations (even results of
molecular simulations, if such are available). We have found that use of the
generalized mean spherical approximation (GMSA) of Waisman \cite{gmsawais},
as implemented by Hoye and Stell \cite{hoye76}, gives considerable
improvement over the original PY approximation. Moreover it still preserves
the analytic simplicity of the resulting $c_{0}(r)$ so that the methods of
Appendix C can be used.

The GMSA approximates $c_{0}(r)$ outside the hard core (we set $d=1$ here),
where PY assumes $c_{0}\ $vanishes, by a Yukawa function: 
\begin{equation}
c_{0}(r>1)=K\frac{e^{-z(r-1)}}{r}.  \label{gmsatail}
\end{equation}
The exact core condition $g(r<1)=0$ then allows one to obtain $c_{0}(r)$
inside the core and satisfy the Ornstein-Zernike equation: 
\begin{equation}
c_{0}(r<1)=-a-br-\frac{\eta a}{2}r^{3}-v\frac{1-e^{-zr}}{zr}+v^{2}\frac{%
\cosh {zr}-1}{2Kz^{2}e^{z}},
\end{equation}
Requiring consistency between compressibility and virial routes for the
pressure and agreement with simulations gives explicit analytic expressions
for $a$, $b$, $v$, $K$ and $z$ as functions of the packing fraction $\eta
\equiv \pi \rho /6,$ as discussed in \cite{hoye76}. We can use this improved
expression for $c_{0}(r_{12};\rho _{0}^{r_{1}})$ in Eq.~(\ref{Wallpy}) to
describe the hard sphere reference fluid.

We can also amend our description of the wall particle in a similar way by
adding the tail correction Eq.~(\ref{gmsatail}) as given by the GMSA to the
right side of Eq.~(\ref{Wallpy}). In the absence of attractive forces this
equation then reduces exactly to a GMSA description of the response of a
hard sphere fluid to a hard sphere fixed at the origin, and we neglect any
changes in this correction when attractions are taken into account through $%
\phi _{R1}(r)$. These GMSA corrections to the usual linear response
treatment are numerically significant only for the high density states
studied in Fig.~6.


\begin{references}
\bibitem{cutoff}  To compare with simulations, we actually used the standard
cutoff and shifted LJ potential with cutoff radius at $r=2.5\sigma $.
\bibitem{preliminary}  A preliminary account of this part of our work was
published in Ref. \cite{wvk}.
\bibitem{iteration}  In practice, since in MD only forces are needed, we
actually iterated the gradient of Eq.~(\ref{mfint}), with forces determined
``on the fly'' during the simulation. This also allowed us to test other
more complicated approximations to the ERF that involved gradients, such as
the ``inhomogeneous force equation'' introduced in Ref. \cite{wvk}. As
already discussed in Ref. \cite{wvk}, only small differences were found, so
we report here only results for the simpler Eq.~(\ref{mfint}).
\bibitem{localdensity}  This is equivalent to the local density
approximation often used in density functional theories. See Ref. \cite
{lit:denfungeneral}.
\end{references}
\end{document}